\documentclass{article}

\usepackage{PRIMEarxiv}

\usepackage[utf8]{inputenc} 
\usepackage[T1]{fontenc}    
\usepackage{hyperref}       
\usepackage{url}            
\usepackage{booktabs}       
\usepackage{amsfonts}       
\usepackage{nicefrac}       
\usepackage{microtype}      
\usepackage{lipsum}
\usepackage{fancyhdr}       
\usepackage{graphicx}       
\graphicspath{{media/}}     
\usepackage{authblk}
\usepackage{float}

\usepackage{amsmath}
\title{Superscreening and polarization control in confined ferroelectric nematic liquids}
\author[1]{Federico Caimi}
\author[1]{Giovanni Nava}
\author[1]{Susanna Fuschetto}
\author[2]{Liana Lucchetti}
\author[3]{Petra Paiè}
\author[4]{Roberto Osellame}
\author[5]{Xi Chen}
\author[5]{Noel A. Clark}
\author[5*]{Matthew Glaser\thanks{matthew.glaser@colorado.edu}}
\author[1*]{Tommaso Bellini\thanks{tommaso.bellini@unimi.it}}

\affil[1]{Dept. of Medical Biotechnology and Translational Medicine, University of Milano, Italy}
\affil[2]{SIMAU Dept., Università Politecnica delle Marche, Ancona, Italy}
\affil[3]{Dipartimento di Fisica, Politecnico di Milano, Milano, Italy}
\affil[4]{Istituto di Fotonica e Nanotecnologie, Consiglio Nazionale delle Ricerche (IFN-CNR), Milano, Italy}
\affil[5]{Dept. of Physics, Soft Materials Research Center, University of Colorado, Boulder, CO, USA}

\begin{document}

\maketitle

\begin{abstract}
The combination of large spontaneous polarization and fluidity
makes the newly discovered ferroelectric nematic liquid crystalline phase (N\textsubscript{F}) responsive to electric fields in ways that have no counterpart in other materials.
We probe this sensitive field response
by confining a N\textsubscript{F} fluid in microchannels that connect electrodes through straight and curved paths. We find that by applying electric fields as low as E $\approx$ 0.5 V/mm, the N\textsubscript{F} phase orders with its polarization smoothly following the winding paths of the channels even when oriented antiparallel to the line
connecting positive to negative electrodes, implying analogous behavior of the electric field. Upon inversion of E, the polar order undergoes a complex multistage switching process dominated by electrostatic interactions. Multistage polarization switching dynamics is also found in numerical simulations of a quasi-2D continuum model of N\textsubscript{F} liquid crystals in microchannels, which also clarify the conditions under which the electric field is guided by the microchannels. Experiments and theory indicate that all observations are direct consequences of the prompt effective screening of electric field components normal to the channel walls.  This electric ``superscreening'' effect emerges as a distinctive property of the N\textsubscript{F} phase, capable of inducing conditions in which both the polarization and the electric field are guided by microchannels.
\end{abstract}


\vspace{30pt}
The discovery of molecules capable of ordering into a ferroelectric nematic (N\textsubscript{F}) liquid crystal (LC), a state predicted a century ago \cite{Weiss1907,Debye1912,Born1916} but only recently observed \cite{Mandle2017,Nishikawa2017,chen2020firstprinciple,Lavrentovich2020}, has attracted immediate interest because of the novel properties of this new fluid state, involving its electric behavior, its elasticity, and its non-linear susceptibilities \cite{Mertelj2018,Chen2021,Sebastian2021,Feng2021,Barboza2022,Sebastian2022}. Conventional nematic LCs are non-polar anisotropic fluids, in which molecules partially align along a common axis, the so-called nematic director $\mathbf{n}$. In the N\textsubscript{F} phase, a bulk electric polarization density $\mathbf{P}$ develops parallel to $\mathbf{n}$
through a weak first order phase transition. Molecules found to give rise to N\textsubscript{F} ordering have large permanent electric dipoles ($> 10$ D) and a high degree of polar orientational order, resulting in a large ferroelectric polarization, up to $P_0 = \lvert \mathbf{P} \rvert \approx 6\ \mu\mathrm{C/cm}^2$ \cite{chen2020firstprinciple,Mandle2017rational,Li2021}.

The presence of a bulk polarization field that is readily reoriented makes the electric field response of (N\textsubscript{F}) markedly different from the dielectric response
of conventional nematics, in ways that have yet to be fully explored. Because of the dielectric anisotropy of conventional nematics, the electric field ($\mathbf{E}$) has a tensorial coupling to $\mathbf{n}$, producing torques on the director proportional to $E^2$, and of a magnitude comparable to the orientational elasticity. In the N\textsubscript{F} phase the presence of a bulk polarization gives rise to a dipolar coupling to $\mathbf{E}$ and self-interaction of space charges arising from divergence of the polarization, producing electric torques orders of magnitude larger. Indeed, because of the fluidity of the N\textsubscript{F} phase, the orientation of $\mathbf{P}$ is a Goldstone variable with energetics dominated by electrostatics, with nematic elasticity and surface anchoring playing a minor role. The unconstrained rotation of $\mathbf{P}$ is affected only by the weak elasticity arising from its non-uniform orientation and by the self-interaction caused by electrostatic forces arising from polarization space charge at surfaces ($\sigma_\mathrm{P} = \mathbf{P\cdot u}$, $\mathbf{u}$ being the unit vector normal to a limiting surface of N\textsubscript{F}) and in the bulk ($\rho_\mathrm{P} = - \nabla \cdot \mathbf{P}$).

All these factors combine to create a fundamentally new kind of fluid, markedly different from conventional nematic LCs and from solid ferroelectric materials, whose polarization, intrinsically constrained by the symmetry of the crystal unit cell, is typically limited to a set of easy axis orientations. In this work we explore the interplay of external fields, geometry, and electrostatic self-interaction N\textsubscript{F} fluids confined within straight and bent microchannels.
Indeed, because of its spontaneous bulk polarization, a ferroelectric nematic fluid confined to a channel should, in the absence of external electric fields and other couplings, align with $\mathbf{n}$ along the channel to minimize the overall electrostatic energy, a behavior with intriguing implications in the case of bent channels. \\

\noindent {\em Description of the experiment} \\

We have produced, via femtosecond laser micromachining assisted by chemical etching \cite{Marcinkevicius:01,osellame2011femtosecond}, microchannels buried in monolithic fused silica that connect gold wire electrodes. The channels, shown in Fig. 1a-b, are all of length $\ell \approx 1\ \mathrm{mm}$ and have a rounded square cross-section of width $w \approx 35\ \mu \mathrm{m}$. Four different shapes are produced: straight along the x-axis direction connecting the electrodes (I-shaped channel), with the central part bent with respect to the x-axis of $90^{\circ}$ (L-shaped channel), $135^{\circ}$ (Z-shaped channel) and $180^{\circ}$ (S-shaped channel) (see Fig. 1b). The channels were filled with RM734, a liquid crystal material that exhibits nematic (N) and ferroelectric nematic (N\textsubscript{F}) phases \cite{Mandle2017,chen2020firstprinciple}, with the N--N\textsubscript{F} transition temperature $T_{\mathrm{N-N}_\mathrm{F}} \approx 133 ^{\circ}C$. Prior to filling, the glass channel surfaces were silanized with hexadecyltrimethoxysilane (HDTMS) to favor planar orientation of RM734 \cite{caimi2021surface}. A detailed description of the fabrication process is given into the SI section.\\

\noindent {\em Response to static fields} \\

In the absence of an applied voltage difference $\Delta V$ between the electrodes, the nematic director exhibits partial ordering along the channel in both the N and N\textsubscript{F} phases (Fig. 1c), probably because of the residual roughness of the channels walls. By applying a voltage $\Delta V$, which gives rise to a “nominal” electric field $E\textsubscript{0} \equiv \Delta V/\ell$, the N\textsubscript{F} ordering becomes virtually perfect in all four channels (Fig. 1c-d) when $E\textsubscript{0} > 0.25\ \mathrm{V/mm}$, while no significant effect is visible in the N phase for fields up to $E\textsubscript{0} \approx 50\ \mathrm{V/mm}$. We closely inspected the field-induced N\textsubscript{F} alignment by polarized transmission optical microscopy (PTOM) and found no sign of either defect lines or defect walls even in the curved sections of the channels that bend contrary to the naïve electrode-to-electrode direction expected for $E_{0}$ in uniform dielectrics (Fig. 1e). This finding indicates that under these conditions both the nematic director $\mathbf{n}$ and the polarization $\mathbf{P}$ are always nearly parallel to the solid surface and continuously follow the channel, as sketched in Fig. 1f.
The continuity of N\textsubscript{F} polarization along the channels also implies a similar continuity of the electric field, which thus follows the channels rather than being directed along the paths connecting the electrodes as in homogeneous dielectric media. This behavior is not dissimilar to what happens in a bent conductive wire between electrodes, where the surface accumulation of free charges steers the field along the wire, independently of its path. In N\textsubscript{F} this same effect must be obtained with bound charges.\\

\begin{figure}[H]
    \centering
    \includegraphics[width=1\textwidth]{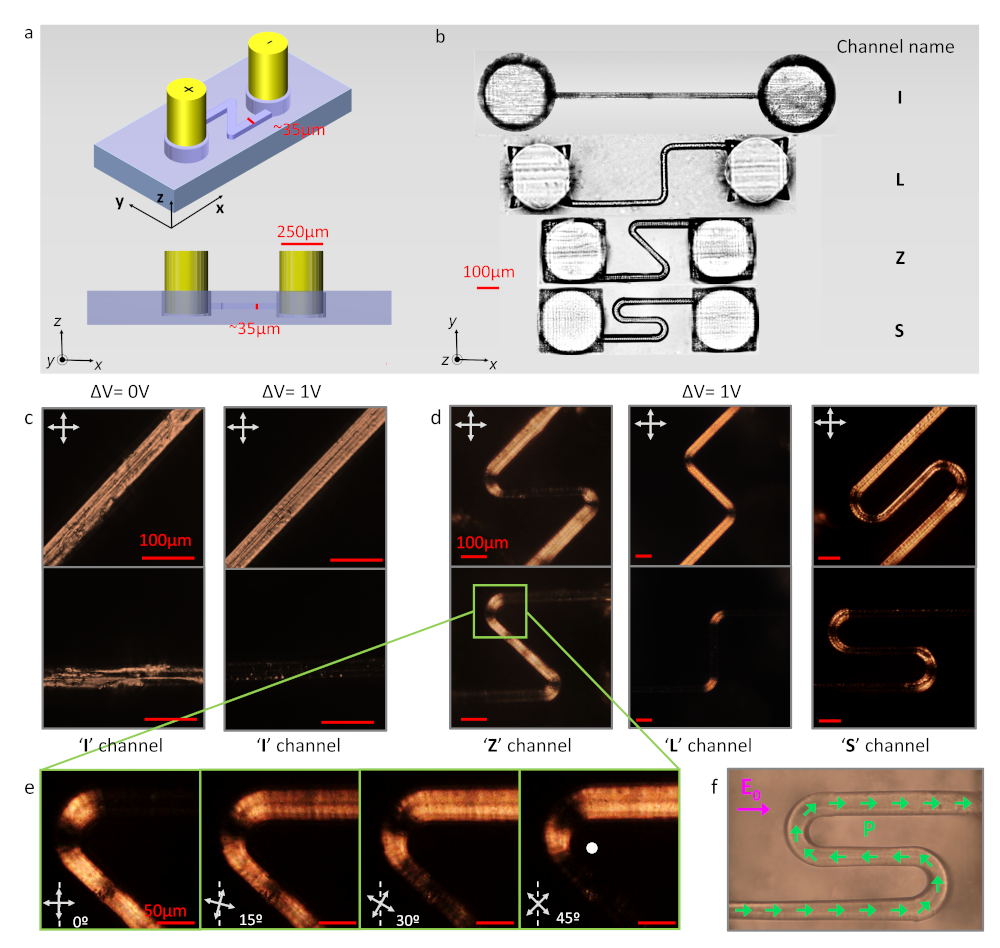}
    \caption{Ferroelectric nematic ordering in microchannels.
a: drawing of the buried microchannels. Yellow cylinders represent the gold wire electrodes. 
b: pictures of the 4 channels considered in this work, designed so that the central part forms an angle of 0°, 90°, 135° and 180° with respect to the x axis.
c: PTOM microscope pictures between cross polarizers in the absence (left) and in presence (right) of an electric field of the “I” channel filled with RM734. The fully dark appearance of the channel in the bottom-right picture indicates that in the presence of a field, the orientation of the N director is along the channel.
d: PTOM microscope pictures between cross polarizers of the Z, L and S channels, respectively, oriented so that the lateral portions are at 45° (top panels) or along (bottom panels) the analyzer while $\Delta V = 1 V$ is applied to the electrodes. 
e: enlargement of a curved portion of the Z channel at various orientations with respect to the polarizer, as indicated.
f: sketch of the continuity of the direction of the polarization $\mathbf{P}$ (green arrows) within the S channel indicating that, in its central section, the polarization is pointing in the opposite direction with respect to the nominal applied electric field $E_{0}$.
}
\end{figure}
\clearpage

\noindent {\em Ferroelectric superscreening} \\

We understand the continuity of polarization and electric field along the channels as the consequence of the orientational freedom of $\mathbf{P}$ in N\textsubscript{F}, enabling an immediate deposition of bound charges that readily cancels the normal components of $\mathbf{E}$ within the channel, which become thus parallel to the channel axis. 

The phenomenon is sketched in Fig. 2. 
The bare (ferroelectric polarization-free) microchannel exposed to a transverse field $E_{0\perp}$ behaves like a capacitor with dielectric constant $\epsilon$ and capacitance per unit length of channel $C$ (Fig. 2a).  In the N\textsubscript{F} phase an additional a bulk polarization field $\mathbf{P}$ is present (Fig. 2b), which has the key features of being reorientable, a result of the symmetry, energy degeneracy, and fluidity of the LC, and spontaneously spatially uniform, to avoid polarization space charge in the LC volume \cite{Zhuang1989}.  $\mathbf{P}$ is drawn parallel to the channel ($\psi$ = 0) since $\psi \neq$ 0 results in the deposition of surface polarization charge/unit length of magnitude $\lambda$ = wP$sin\psi$ on the channel edges (Fig. 2c).  The resulting electric field $E_P(\psi)$ within the channel provides a restoring torque density that opposes the reorientation, a sort of electrostatic orientational spring that tends to force $\psi$ back to zero.  Application of an external $\psi$-independent torque results in a balance with this restoring torque at finite psi (Fig. 2c).  This electrostatic spring is exceptionally stiff, as can be estimated from $V_{sat} = Pw/C$, the voltage generated across the channel when the orientation is saturated at $\psi = 90^\circ$ (Fig. 2d.). For our channels we estimate $V_{sat} \approx 10^6$V so that $\mathbf{P}$ being perpendicular to the channel is an idealized condition actually never achievable in practice since it would require creating a gigantic electric field. 
 
In the channel experiments reported here the external torque is produced by an applied electric field $\mathbf{E}_{0}$ forming an angle $\beta$ with the channel axis (Fig. 2e), giving rise to a $\mathbf{P}$ rotation and restoring torque as in Fig. 2c. In this case the torque balance condition is $E_{\perp}  = 0$ within the channel, where $E_P(\psi)$ cancels $E_{0\perp}$ \cite{Clark2000,Shen2011,Coleman2003}. For typical experimental $\mathbf{E}_{0}$ magnitudes and a typical N\textsubscript{F} polarization $P \approx 6 \mu C / cm^2$,  this cancellation occurs for sub-microradian reorientations of $\mathbf{P}$, with $\psi = w E_{0\perp}/ V_{sat}$,  leaving $\mathbf{P}$ always essentially parallel to the local channel axis, $E_{\parallel} = E_{0\parallel}$ and $E_{\perp}$  = 0 in the channel. Such immediate and effective screening, by which the normal component of $E_0$ is readily cancelled, is a distinctive property of the N\textsubscript{F} phase that we call {\em ferroelectric superscreening} to mark the difference between this ultimately collective fluid dipolar reorientation and conventional dielectric or conductive screening by dipoles or free charges.  

In the microchannel experiments, ferroelectric superscreening locally cancels the normal electric field component, but also globally affects its longitudinal components through the long-range effect of the modified bound charge distribution.\\

\begin{figure}[t]
    \centering
    \includegraphics[width=1\textwidth]{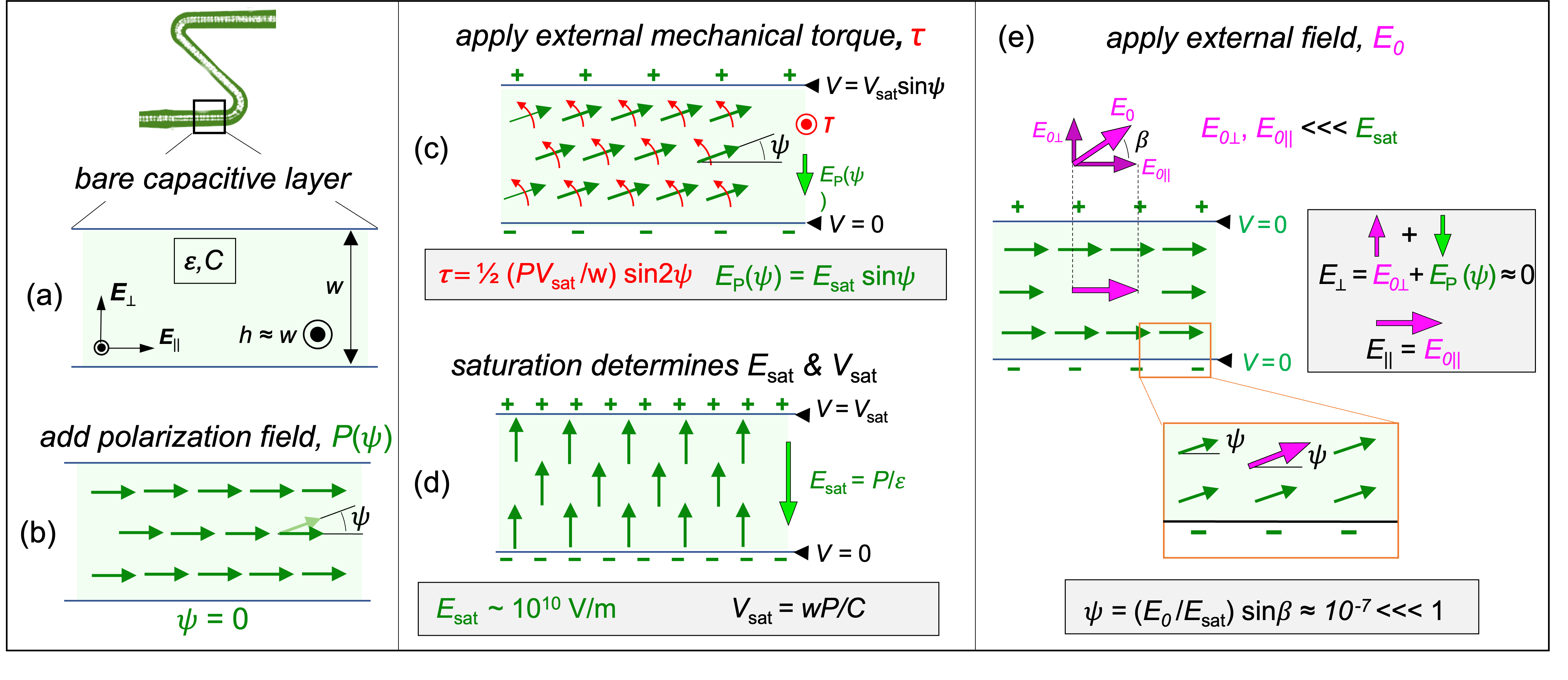}
    \caption{Schematic description of superscreening in a channel.  Schematically, the N\textsubscript{F}-containing channel can be considered a dielectric slab of dielectric coefficient $\epsilon$ into which a reorientable polarization field $\mathbf{P}(r)$ has been introduced and then exposed to external torques.  (a) The bare channel has a capacitance/length of channel $C$, which is the ratio of voltage $V$ across the channel to $\lambda$, the charge along each edge per unit length of channel, such that $\lambda = CV$. (b)  Being the reorientable field $\mathbf{P}$ locally along the nematic director, its spatial variation is controlled by nematic elasticity and by the coupling of P to electric fields, both external and internal, due to polarization space charge and polarization surface charge. The ratio $\sqrt{\epsilon K}/P$, where K is a Frank elastic constant, is very small (less than $1$ nm), indicating that electrostatics is dominant, a condition that makes $\mathbf{P}$ uniform and its orientation $\psi$ responsive to external torques as a spatially uniform field. (c) application of an external $\psi$-independent mechanical torque is balanced by an electrostatic restoring torque per unit length of channel $\tau = (P^2/C) sin\psi$ due to polarization surface charge at the channel edges.  (d) If $\mathbf{P}$ is reoriented to be normal to the plates, then $V \equiv V_{sat} = P/C$, and the orientational potential energy/length will be $P^2/C = PV_{sat}$.  The corresponding electric field in the channel midplane will be $E_{sat} \sim 10^{10} V/m$, $i.e.$ $V_{sat}\sim10^5 V$, many orders of magnitude larger that the fields applied in the experiments. (e) microchannels in the experiments are in electric fields $\mathbf{E}_0$ not generally oriented with the channel axis, with components $E_{0\parallel}$ and $E_{0\perp}$ along and perpendicular to the channel axis, respectively. While $E_{0\parallel}$ is continuous across the channel walls, $E_{0\perp}$ is cancelled by superscreening. Being the applied fields much smaller than $E_{sat}$, the tilt angle $\psi$ is so small not to be detectable (in the zoom panel $\psi$ is enhanced to enable representation). }
\end{figure}

\noindent {\em Field reversal experiments} \\

To further explore the role of bound-charge screening in the `polarization following', we investigated the response to the reversal of the sign of $\Delta V$. Upon switching $\Delta V$, the nematic ordering becomes suddenly unstable, with the transmitted light intensity through crossed polarizers ($I_P$) and without analyzer ($I_S$) transiently changing during a time interval that depends on $\Delta V$, being of the order of 100 ms for $E_0 = 1 V/mm$. Specifically, we have measured $I_P$ vs. time in portions of the channels oriented as the polarizer (thus dark in stationary condition as in Fig. 1c lower right panel), and normalized it with $I_{P_{45}}$, the value of $I_P$ measured  in equilibrium conditions (i.e. in stationary state with applied field) with the channel at $45^{\circ}$ with the polarizer (Fig. 1c upper right panel). Analogously, we normalized $I_S$ with $I_{S_0}$, the transmitted intensity in equilibrium conditions. By inspecting the two transmittances $\tau_P(t) = I_P(t)/I_{P_{45}}$ and $\tau_S(t) = I_S(t)/I_{S_0}$ and associated videos, we can recognize three phases for the polar switching in the channels: a disordering regime, a reorganization regime, and a defect-annealing regime. Because of the small section of the microchannels we couldn't measure experimentally the polarization current flowing through the electrodes and compare it to $\tau_P(t)$ and $\tau_S(t)$. We instead performed an analogous measurement on a flat cell (see SI). \\

Inspection of PTOM pictures and videos shows that, upon potential inversion ($t = 0$), the order is replaced by an irregular pattern formed by micron-sized domains (Fig. 3a-b) coherent with the notion that $\mathbf{P}$ locally rotates away from the original orientation in a variety of directions. This leads to an increase in $\tau_P(t)$ due to rotated optical axis (Fig. 3 d-g) and in a decrease in $\tau_S(t)$ due to scattering (see SI). The disorder in the N\textsubscript{F} structure grows to a maximum and decreases, as shown by appearance of a minimum in $\tau_S(t)$ (see SI) found at about the same time $t \approx t_m$ at which $\tau_P(t)$ shows a dent in its growth. 

We find that $t_m$ marks the transition between two different regimes. Pulsed field reversals, in which the sign of $ \Delta V$ is inverted for a time $t_{inv} < t_m$, give rise to a simple exponential relaxation of $\tau_P(t)$, while for longer pulses ($t_{inv} > t_m$), the relaxation process in the channel becomes much longer and irregular (see SI). We thus interpret $t_m$ as the time it takes to the N\textsubscript{F} domains to rotate enough to start reorganizing in the opposite direction, merging and reducing defect lines and lowering the turbidity. 

In the reorganization regime ($t > t_m$), $\tau_P(t)$ still grows, reflecting coarsening and smoothing of the defects, to reach a maximum and decrease back to $0$ when the uniform polar ordering is established in the reversed direction. Such uniform N\textsubscript{F} ordering develops through a nucleation process taking place in the narrowest section of the channel, typically (but not always) located halfway between electrodes as a natural result of the etching process or due to design (see SI). This final stage in the reversal of $\mathbf{P}$ requires major rearrangements of the director that involve disentangling of topological defects, closure of defect loops and reversal of the polarity at the surface. Previous observations in silanized flat cells suggest the latter to be the slowest component of switching after field reversal \cite{caimi2021surface}.

After its nucleation, the defectless uniform region ($\tau_P = 0$) progressively expands via the motion of the interface which separates defected and defect-free N\textsubscript{F} phase (Fig. 3b, bottom panels, and Fig. 3c). Thus, the completion of the $\mathbf{P}$ reversal takes place at a time that grows with the distance from the nucleation site, sooner in proximity of the channel center (Fig. 3d, red line) and later (even after up to 1 s) close to the electrodes (Fig. 3d, blue and green lines).

\begin{figure}[t]
    \centering
    \includegraphics[width=1\textwidth]{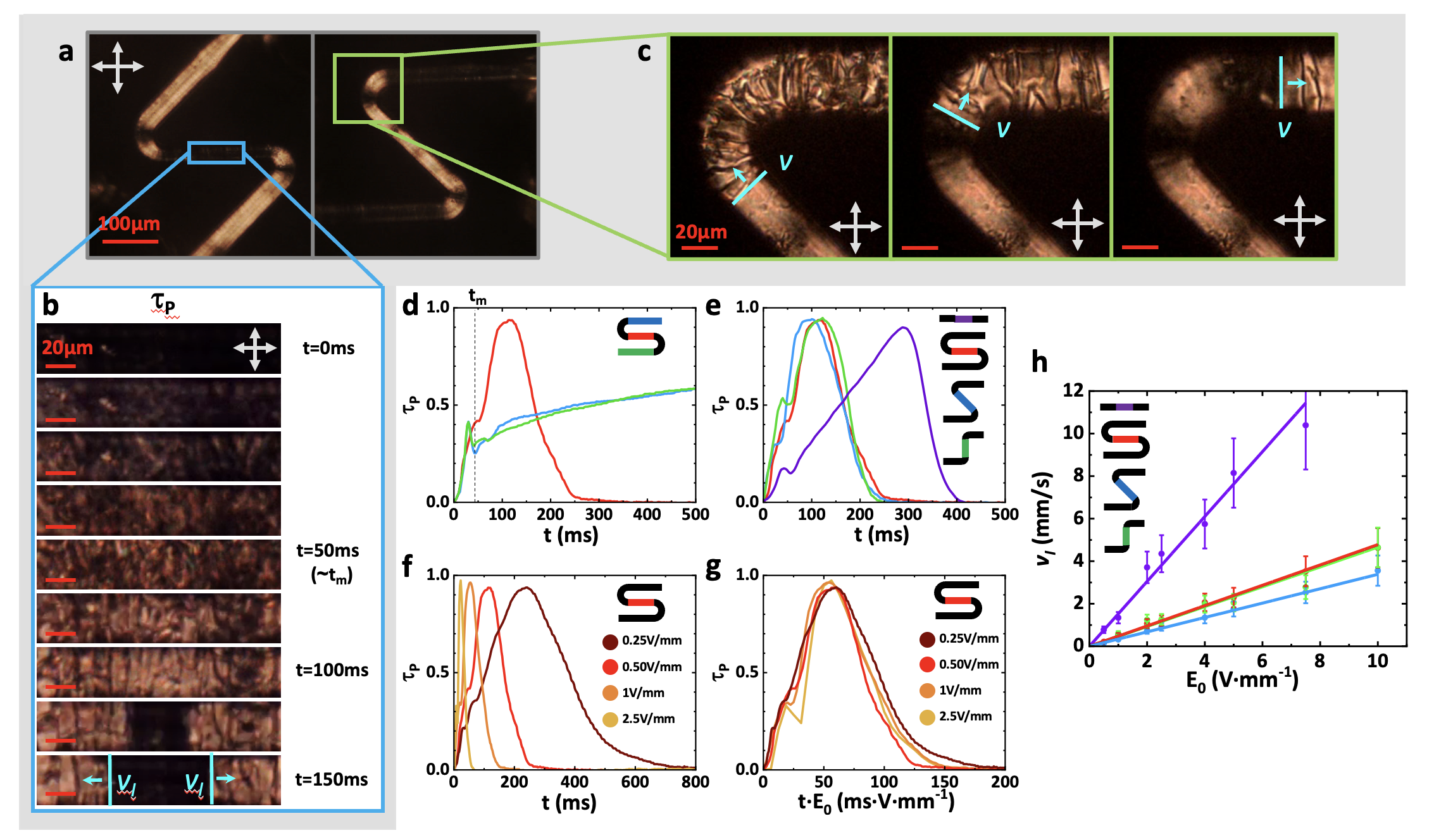}
    \caption{Polarization inversion in the microchannels. a: PTOM microscope pictures of the Z channel with two different orientations with $\Delta V$ = 1V.
b: time sequence of PTOM images of the central part of the Z channel following the field inversion (t=0). The sequence shows the appearance and coarsening of domains and the formation of the homogeneous-defected interface, that moves toward the electrodes with velocity $v_I$.
c: time sequence of PTOM images of a curve  of the Z channel showing the motion of the homogeneous-defected interface with velocity $v_I$.
d: time evolution of the depolarized transmittance $\tau_P(t)$ measured in the central part and in the two side arms of an S-channel during the inversion of $\Delta V$ from 0.5V to -0.5V.
e: time evolution of the depolarized transmittance $\tau_P(t)$ measured in the central part of the various types of channels during the inversion of $\Delta V$ from 0.5V to -0.5V.
f: time evolution of the depolarized transmittance $\tau_P(t)$ measured in the central part of an S-channel during the inversion of different $\Delta V$ between the electrodes resulting in different $E_0$, as in the legend. 
g: Same data of the previous graph plotted on the rescaled time t·$E_0$.
h: Velocity $v_I$ of the homogeneous-defected interface vs the the nominal electric field $E_0$ measured in the central part of the various channels in the first instants after its formation.}
\end{figure}

New insights emerge from the comparison of $\tau_P(t)$ in different positions within the same channel and in different channels. First, the short time behavior is the same everywhere in all positions of  bent channels. This is shown in Fig. 3d for the S channel, while very analogous plots are obtained for the L and Z channels. Second, $\tau_P(t)$ measured in the central part of the bent channels, i.e. close to nucleation site, is basically equal in all regimes provided the applied voltage is the same, as shown in Fig. 3e for $\Delta V = 0.5 V$. Remarkably, $\tau_P(t)$ is instead different for the simple straight geometry, which displays a slower rise and a steeper decay.\\

Fig. 3f shows that the polarization reversal is a process fully controlled by electrical interactions: the evolution of $\tau_P(t)$ becomes faster for larger $\Delta V$ (and hence $E_0$) with its whole kinetics being proportional to $E_0$ in all regimes, as shown in the collapse in Fig. 3g. The dent at $\tau_P(t_m)$ is more pronounced with larger $E_0$, a condition in which we also find a larger turbidity (see SI). This behavior reflects the fact that faster switching occurs through the breaking up of the system in smaller domains, leading to a larger scattering of light.

By demonstrating that the polarity inversion linearly depends on $E_0$, the results in Fig. 3f-g convey the notion that $\tau_P(t)$ can be used as a proxy for the local field $E_L$. Accordingly, the identical rise of $\tau_P(t)$ for $t < t_m$ in the various portions of the channels indicates that, in that time interval, $E_L$ is uniform throughout the channels. Moreover, the identical full shape of $\tau_P(t)$ in the central part of the L, Z and S channels reveals that $E_L$ is the same in the central portion of all bent channels, provided that width, $\Delta V$ and $\ell$ are the same. \\

In the latest stage of the switching, defect-free N\textsubscript{F} domains nucleate in the slimmest portions of the channels, suggesting that those are the places where $E_L$ is largest, and expand via a moving interface. We measured the propagation velocity $v_I$ of such interface for different $\Delta V$ in a limited region closest to the nucleation spot. We find $v_I$ to be constant, and thus well-defined, at least in the first $100 \mu m$ of displacement, away from electrodes and channel bends (see SI). In Fig. 3h we plot $v_I$ vs. $\Delta V$ for the various channels. We find $v_I$ to be approximately linear in $\Delta V$ and to be very similar in all the bent channels. Again, the straight channel makes an exception, with a larger, about double, $v_I$. 

The linear dependence of $v_I$ on $\Delta V$ indicates that the motion of the interface is electrically driven and follows the local electric field, confirming that the latter is directed from electrode to electrode along the channels, despite their bends. Such motion also indicates that the interface carries electric charge. This is coherent with its marking a discontinuity in polarization between the fully polarized uniform phase and the less polarized defected structure, a concept that we further validated through measurements of polarization current and transmittance in flat cells (see SI). The constant  $v_I$ for each $\Delta V$ reveals that the interface motion is opposed by a viscous-type friction force, quite likely reflecting the viscosity arising from rotation of the domains, necessary to free the volume from defects (see SI). \\

\noindent {\em Continuum modeling of switching and equilibrium structure} \\

To gain insight into the experimentally observed phenomena, we performed numerical simulations of the equilibrium structure and switching dynamics of N\textsubscript{F} fluids confined to microchannels.
We assume that the ferroelectric polarization $\mathbf{P}$ has fixed magnitude $P_0$ but can vary in direction, i.e., $\mathbf{P} = \mathbf{P}(\mathbf{r}) = P_0 \mathbf{p}(\mathbf{r})$, where the polar director $\mathbf{p}$ is a unit vector, assumed to be parallel to the nematic director $\mathbf{n}$. We further assume that the $z$ dimension of the channels is large relative to the $x$ and $y$ dimensions, and that all quantities ($\mathbf{n}$, $\mathbf{P}$, $\mathbf{E}$, $\rho$, etc.) are independent of $z$, e.g., $\mathbf{P}(\mathbf{r}) = \mathbf{P}(x,y)$. Finally, we assume that $\mathbf{P}$ and $\mathbf{E}$ are confined to the $x$-$y$ plane, i.e., that $E_z = 0$ and $P_z = 0$. These simplifying assumptions result in a computationally tractable quasi-2D model that captures the essential physics governing the behavior of N\textsubscript{F} fluids in microchannels.

The free energy density of the ferroelectric nematic phase is taken to be the sum of orientational elastic and electrostatic contributions, $f(x,y) = f_\mathrm{elast}(x,y) + f_\mathrm{elect}(x,y)$. The former is the Frank-Oseen orientational elastic free energy density in the one-elastic-constant approximation, $f_\mathrm{elast} = \frac{1}{2} K \lvert \nabla \theta \rvert^2$, where $K$ is the elastic constant for splay and bend deformations and $\theta(x,y)$ is the orientation of $\mathbf{p}$ in the $x$-$y$ plane. 
The electrostatic free energy density is
$f_\mathrm{elect} = - \mathbf{P} \cdot \mathbf{E} = - P_0 \left( E_x \cos \theta + E_y \sin \theta \right)$,
where $\mathbf{E}$ is the total electric field arising from both bound and free charges.

The equilibrium structure is obtained by functional minimization of the total free energy, which we accomplish using a relaxation method: the system is evolved to obtain the equilibrium  $\theta(x,y)$ by solving the equations of motion that result from setting the total torque (the sum of elastic, electrostatic, and viscous torques) to zero.  

The elastic, electric and viscous torques acting on $\theta(x,y)$ are 
$\tau_\mathrm{elast}(x,y) = K \nabla^2 \theta$,
$\tau_\mathrm{elect}(x,y) = \left( \mathbf{P} \times \mathbf{E} \right) \cdot \mathbf{z} = P_0 \left( E_y \cos \theta - E_x \sin \theta \right)$,
and 
$\tau_\mathrm{visc}(x,y) = - \gamma \partial \theta / \partial t$,
where $\gamma$ is the rotational viscosity.
Starting from a given initial condition, the equations of motion are evolved to reach equilibrium  ($\tau_\mathrm{elastic} + \tau_\mathrm{elect} = 0$), and the same equations of motion are solved to model the switching dynamics upon field reversal. These equations are solved numerically, with $\theta$ defined on a discrete grid with grid spacing $a$, using an explicit Euler time-stepping scheme. The geometry of the simulated channels (S, L and I)  matches that of the real channels in the $x$-$y$ plane. For computational expediency, we perform simulations with a potential difference between the end of the channels of $+1000$ V, $i.e.$ about $10^3$ times larger than that applied in the experiments, since with this value the entire switching process is largely complete within $500\ \mu$s. A grid spacing of $a = 1.5\ \mu$m was used in the calculations presented here, but we have verified that varying grid spacings give similar results (see SI). In the simulations described below, we have also introduced homogeneous surface roughness of amplitude $\delta = 0.76\ \mu$m into the channel boundaries, to more faithfully model the experimental situation (see the SI for a discussion of the effects of surface roughness on equilibrium structure and switching dynamics). Full details of the computational method are provided in the Methods section and in the SI. \\

\begin{figure}[t]
    \centering
    \includegraphics[width=1\textwidth]{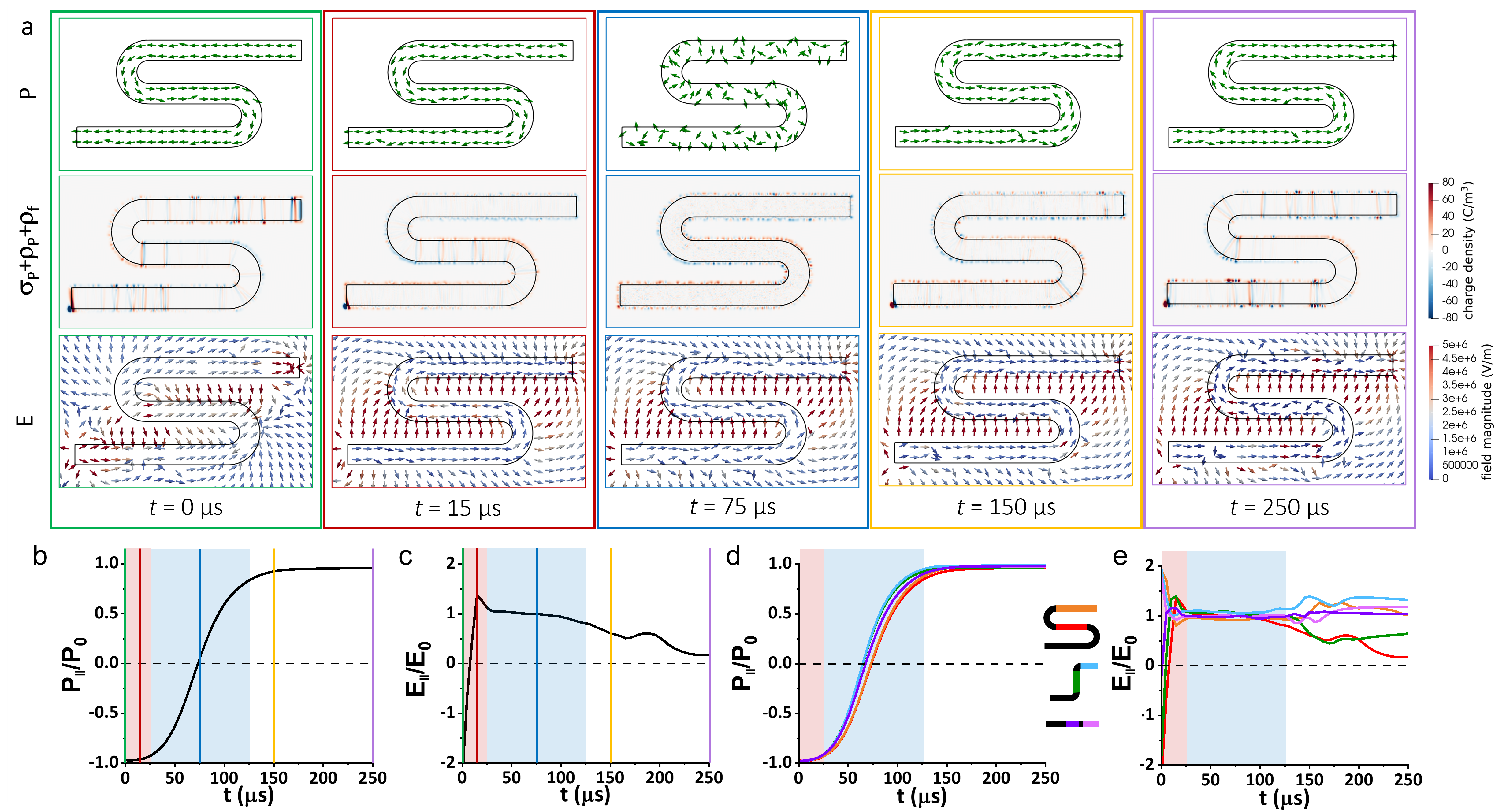}
\caption{ Simulated time evolution of polarization $\mathbf{P}$, total (free + bound) charge density $\rho = \sigma_P + \rho_P + \rho_f$, and electric field $\mathbf{E}$, following sign reversal of the potential on the electrodes (at $t=0=t_0$). The starting condition is the equilibrium configuration for the opposite sign of potential difference. All channels are of width $40\ \mu$m and contour length $\ell \approx 1$ mm. 
(a): Snapshots showing $\mathbf{P}$, $\rho$ and $\mathbf{E}$ in the S-shaped channel at five specific times marked by color frames. The arrows denoting $\mathbf{P}$ and $\mathbf{E}$ are of fixed size, with the field amplitude indicated by the color of the arrows. For clarity, $\mathbf{P}$ and $\mathbf{E}$ are shown only at a subset of grid points (1 in 196).
(b) and (c): $P_{\parallel}(t)$ and $E_{\parallel}(t)$ averaged over the central section of an S-shaped channel. Vertical colored lines mark the times relative to the snapshots in panel a of matching color. (d) and (e): $P_{\parallel}(t)$ and $E_{\parallel}(t)$ from different regions of S-, L-, and I-shaped channels, with the color code indicated in the schematic diagram. The color shading in panels (b)-(e) highlights the time intervals during which the electric field reverses (red shading), and during which the polarization reverses and $E_{\parallel} \approx 1 V/\mu m = \Delta V/\ell$, indicating electric field guiding (blue shading). }
\end{figure}

\noindent {\em Simulation results} \\

The general features of the equilibrium structure and switching dynamics of ferroelectric nematics confined to microchannels can be understood from the simulated switching dynamics upon voltage reversal for the S-shaped channel, shown in Fig. 4. Fig. 4a shows the time evolution of polarization density, total charge density, and electric field upon voltage reversal. Figs. 4b and 4c show the $t$ dependence of $P_{\parallel}$ and $E_{\parallel}$, the components of $\mathbf{P}$ and $\mathbf{E}$ parallel to the local channel direction, averaged over the central linear section of the S-shaped channel, where the colored vertical lines indicate the times corresponding to the snapshots in Fig. 4a with matching colored frames. Immediately after voltage reversal ($t = 0$) the electric field in the channel is nonuniform in magnitude and has a substantial component transverse to the channel. This transverse component of field is rapidly screened out by a slight reorientation of the polarization at the channel surfaces, and by $t = 15\ \mu$s (red vertical line) the electric field in the channel follows the channel contour and is nearly uniform in magnitude ($\sim 1\ \mathrm{V}/\mu$m) (see Fig. 3e, red frame). In this time interval, the field outside the channel switches direction, with a corresponding change in sign of the bound charge density at the channel surfaces. Fig. 4d-e shows that the same field-reversal time interval during which $\mathbf{E}$ changes direction while $\mathbf{P}$ remains nearly constant is observed in different regions within the L, S and I channels, identified by the color code in the schematic diagram between Fig. 4d and 4e. 

At the end of the field-reversal process, the polarization field in the channel is antiparallel to the electric field and thus in a state of unstable equilibrium, leading to reorientation and disordering of the polarization field at later times. The configuration at time $t = 75\ \mu s$ (corresponding to $P_\parallel(t_2) \approx 0$) is shown in the blue-framed snapshots of Fig. 3e, where it appears that the inversion of $P_\parallel$ takes place through a disordered state, indicating that reduction of the free energy occurs via breaking of the polar director field into small domains, enabling local minimization of the energy cost due to bulk and surface charge accumulation. The orientation of the electric field within the channel becomes somewhat disordered during this process, but overall $E_\parallel$ and $\rho$ remain nearly constant in the interval $t = 25$--$125\ \mu s$ (blue shading of Fig 4b-e), the time interval during which $P_\parallel$ nearly completes its inversion. Thus, most of the $\mathbf{P}$ inversion process takes place while $E_\parallel$ is roughly constant and uniform. Very similar behavior of $P_\parallel$ and $E_\parallel$ is found over this time interval for different regions of the various channels (Fig. 4d-e).

Polarization reversal is not the final event in the voltage reversal response. Indeed, a complex subsequent dynamics takes place, involving a more subtle reorientation of the polarization field and an associated reorganization of bound charge in bands of positive and negative sign transverse to the channel, resulting in nonuniformity in the magnitude of the electric field along the channel, evident by $t = 250\ \mu$s (Fig. 4a, purple frame). In this last stage, $E_{\parallel}(t)$ evolves differently in different regions of the various channels (see Fig. 4e for $t > 125\ \mu$s) and depend sensitively on details of channel roughness, as can be seen by comparison with simulations in channels in which the linear portions of the channels are smooth (see data and discussion in SI), making impossible to identify an equilibrium state to compare with experiments.

Switching in the three simulated microchannels thus takes place in three distinct stages: (i) fast screening of the normal component of the electric field at the boundaries of the channel (`superscreening’) leading to uniform following of the channel contour by the electric field, and accompanied by reversal of the field direction outside the channel and an associated change in the sign of bound charge at the surface of the channel; (ii) reversal of the polarization direction via a disordered intermediate state, culminating in uniform following of the channel contour by the polarization field while $\mathbf{E}$ within the channel remains approximately uniform in magnitude and guided; (iii) roughness-dependent reorganization (`annealing’) of the polarization field and associated bound charges within the channel.

Finally, to clarify the role of channel width variation, we performed test simulations on S and I channels tapered in the central region, finding that indeed polarization reversal is faster and the local field larger where the channel has the minimum width (see SI).  \\

\noindent {\em Discussion} \\

Despite the strong simplifications adopted in the computation, the simulation results clearly confirm that the major experimental observations are direct consequence of the unique features of the N\textsubscript{F} fluid state. The polarization continuously follows the channels from electrode to electrode even in bent-shape geometries, in full agreement with the experimental observations.

The simulations confirm the essential role of N\textsubscript{F} superscreening, through the prompt accumulation of surface polarization charges that screen the component of $\mathbf{E}$ normal to the channel boundaries, causing the electric field inside each channel to become directed along the channel, with a magnitude that, during the $\mathbf{P}$ switching, is roughly constant and uniform in all channels. This notion agrees with the observation that the electrically controlled switching kinetics (initial part) is indeed similar in the different positions of the bent channels.

Simulations also confirm that the $\mathbf{P}$ switching taking place through the observed disruption of the nematic ordering into small domains is also a consequence of the bulk polarization, a strategy by which the system minimizes the surface and bulk bound charge during inversion. Of course, missing the third dimension, this result remains a qualitative indication only. 

On longer timescales, simulation predicts variations in electric field magnitude that depend on the detailed shape of the channel. Experiments indicate that in the latest stages of $\mathbf{P}$ inversion, the defect-free volume nucleates in the narrowest sections of the channels where simulation suggest the field is largest, while the expansion velocity of such domains ($v_I$) is equal in the bent channels, a sign that the electric field has remained equal in the central part of the S, Z and L channels. To assess the quantitative coherence of these observations a much more detailed level of modeling would be required. 

Experiments show that straight channels behave a bit differently than the bent ones, with a slower onset of $\mathbf{P}$ disordering and a larger $v_I$. This is not in simple agreement with simulations, which predict a generally uniform behavior. Even in simulation, however, the straight channel has a singular behavior: when built with smooth walls shows a much slower $\mathbf{P}$ inversion; tapering and/or introducing wall roughness has a more dramatic effect than in bent channels. We understand the peculiarity of the straight channel as a consequence of its symmetry, by which no easy point of charge accumulation are available, as in the bends of the other channels. Elimination of the normal component of $\mathbf{E}$ requires in the straight channel the buildup of a nonzero $\nabla \cdot \mathbf{P}$, while bending, tapering and wall noise provide immediate accumulation points for polarization charges, which speed up the charge distribution process. We thus speculate that the quantitative features of the straight channel switching behavior depend on its construction details.  \\
\clearpage
\noindent {\em Conclusions} \\

The electric field response of ferroelectric fluid material confined in bent microchannels, by which  polarization and electric field follow the channels along their paths, brings to light the capacity of these materials of promptly cancelling any electric field components normal to dielectric surfaces. 
This unique and defining property of N\textsubscript{F}, which we have called `superscreening', is achieved through minor local changes in the orientation of $\mathbf{P}$ and is the leading and fastest component of the multiscale dynamic response of N\textsubscript{F} to electric fields, determining all the following evolution. 

Superscreening mandates that the electric field within the channel `follows’ the channel, which further implies that the ferroelectric polarization field follows the contour of the channel under equilibrium conditions because the electrostatic energy is minimized when the polarization field is locally parallel to the electric field everywhere. This behavior strictly depends on the electrostatic energy being the dominant free energy contribution (in our conditions about $10^4$ times larger than the elastic energy, see SI), a feature making ferroelectric nematics strikingly different to conventional nonpolar nematics.  

The S-shaped channel, in which there are three parallel three straight sections, is particularly revealing: in principle, the polarization field could point in the same direction (say to the right) in all three sections, with polarization reversal walls in the two curved portions of the channel. However, such polarization reversal walls are highly charged and thus inherently unstable, their bound surface charge density $\sim \pm 2 P_0$ giving rise to fields on the order of $10^{10}$ V/m. Should such reversal walls transiently form, the large electric field they generate would immediately reorient the polarization in the central portion of the S-shaped channel to produce uniform `polarization following’ along the channel contour, as observed.

Superscreening operates at the N\textsubscript{F}-dielectric interface, but not when N\textsubscript{F} material is in direct, high capacitance contact with electrodes, such as those at the ends of the channels in the experiments described here, where bound polarization charges are effectively compensated by free charges, a condition enabling strong electro-optical response, as the one following the potential reversal. 

The last step in the switching involves annealing of defects in the bulk and polarity inversion at the surfaces. This occurs through a nucleation process that starts where the field is largest and propagates through the entire channel via an electrophoretic-type motion of the interface between uniform and disordered regions. We find the whole behavior to be electrically driven, and thus specific to this fluid ferroelectric phase in which electrostatic interactions dominate.

Our findings open to the possibility of exploiting the propagation of order within channels to design microconfinement geometries that provide a simple means to control the optical axis in multiple positions in electro-optic devices. More broadly, this study establishes a conceptual framework for understanding the behavior of N\textsubscript{F} materials in a variety of settings in which geometrical confinement effects are relevant, e.g., in composite structures, in porous and/or disordered media with quenched or annealed disorder, and in the presence of topographically patterned substrates.

\section*{Materials and methods}

\noindent {\em Materials} \\
RM734 (4-[(4-nitrophenoxy)carbonyl]phenyl 2,4-dimethoxybenzoate) has been synthesized as reported in \cite{chen2020firstprinciple}, all the other reagents and materials were obtained from different commercial sources.\\

\noindent {\em Channel fabrication} 

Microchannels have been fabricated by femtosecond laser irradiation followed by chemical etching (FLICE) \cite{Marcinkevicius:01,osellame2011femtosecond} in 1 mm thick fused silica substrates. Details of the whole process are provided in the Supplementary Information. Following laser irradiation, we have exposed the substrate to two different etchant solutions: first an aqueous solution of hydrofluoric acid (HF) to etch the access holes, and then a solution of potassium hydroxide (KOH) to etch the channels. We have used this dual process, to benefit from the advantages of both the etchants. Indeed, HF guarantees high etching rate for large volume removal, i.e. for the creation of electrode access holes, while KOH is slow but provides high etching selectivity for controlled and uniform microchannel cross-section over the entire length \cite{LoTurco_2013}. 
To smooth the microchannel surfaces, we annealed the samples in a temperature-controlled furnace \cite{sala2021effects}. Channel surfaces have then been treated using a modified version of the procedure reported in \cite{caimi2021surface}. \\

\noindent {\em Transmittance measurements} \\

Transmitted light intensities trough crossed polarizers ($I_P$, $I_{P_45}$) and without analyzer ($I_S$, $I_{P_45}$) are measured recording 400FPS videos with a CMOS microscope camera. Analysis are performed on a $30\times30\mu m^2$ region, aligned along the polarizers (thus dark in stationary conditions) and centred on the interface nucleation point for measurements for the central branches and at halfway for the lateral branches. Transmittance trough crossed polarizers $\tau_P (t)$ is calculated as the ratio between $I_P (t)$ and the intensity measured in stationary condition at $45^\circ$ with the polarizer. Transmittance without analyzer $\tau_S(t)$ is calculated as the ratio between $I_S(t)$ and the intensity measured in stationary condition at the instant $t_0$ ($I_{S_0}$). All the measurements are obtained as an average of at least three negative to positive switches and three positive to negative switches. \\

\noindent {\em $v_i$ measurements} \\

To perform $v_i$ measurements 500FPS videos are recorded with a CMOS microscope camera and analyzed with a tracking-edge MATLAB script whose functioning is better described in S.I.. Measurements are referred to a single interface and recorded in the first instants after its formation. All the measurements are obtained as an average of at least two negative to positive switches and two positive to negative switches. Error bars are calculated considering both errors on edge position extimation and averaging between switches.\\

\noindent {\em Simulations} \\

The dynamical evolution and equilibrium structure observed in our simulations depend on initial conditions and other details of the simulations, and we have carried out a variety of numerical experiments to assess these dependencies.
For the results presented here, we adopt the following simulation protocol, designed to minimize the dependence on initial conditions: (1) initialize the system with a completely random polar director field (the polar angle $\theta$ at each grid point is assigned a random value in the range $-\pi, \pi$); (2) evolve the system forward in time for $510\ \mu$s while maintaining a constant potential difference of $-1000$ V between the two ends of the channel; (3) change the sign of the voltage between the ends of the channel to $+1000$ V and evolve the system forward in time for $500\ \mu$s. Here, we present results from stage (3) of these simulations, under the assumption that this models the experimental switching dynamics at a qualitative level, and that the final configuration is a reasonable model of the equilibrium state of the system for a fixed applied voltage. In our model, the free charge on the electrodes is represented by a uniform charge density located at and spanning the channel ends, with equal and opposite magnitude at the two ends of the channel. A constant electrostatic potential difference $\Delta V$ between the two ends of the channel is maintained by appropriately scaling the free charge density at each timestep. Full details of the theoretical model and computational methods can be found in the SI. 

\bibliographystyle{unsrt}  
\bibliography{references}

\end{document}